\documentclass[11pt]{article}

\newcommand{\bmat}{\left(\begin{array}}
\newcommand{\emat}{\end{array}\right)}

\def\yzero{\smash{\hbox{$y\kern-4pt\raise1pt\hbox{${}^\circ$}$}}}

\def\beq{\begin{equation}}
\def\eeq{\end{equation}}
\def\beqa{\begin{eqnarray}}
\def\eeqa{\end{eqnarray}}

\def\-{\hphantom{-}}

\def\s2{\frac{1}{\sqrt2}}

\def\beq{\begin{equation}}
\def\eeq{\end{equation}}
\def\beqa{\begin{eqnarray}}
\def\eeqa{\end{eqnarray}}

\def\IF{\relax{\rm I\kern-.18em F}}
\def\II{\relax{\rm I\kern-.18em I}}
\def\IP{\relax{\rm I\kern-.18em P}}
\def\IC{\relax\hbox{\kern.25em$\inbar\kern-.3em{\rm C}$}}
\def\IR{\relax{\rm I\kern-.18em R}}

\def\cp{{\cal P}}

\def\Dsl{\,\raise.15ex\hbox{/}\mkern-13.5mu D} 
\def\IZ{Z\kern-.4em  Z}

 \def\cp#1{\relax\ifmmode {\IP\kern-2pt{}_{#1}}\else $\IP\kern-2pt{}_{#1}$\=fi}

\catcode`\@=11
\newdimen\@rotdimen
\newbox\@rotbox

\def\@vspec#1{\special{ps:#1}}
\def\@rotstart#1{\@vspec{gsave currentpoint currentpoint translate
   #1 neg exch neg exch translate}}
\def\@rotfinish{\@vspec{currentpoint grestore moveto}}
\def\@rotr#1{\@rotdimen=\ht#1\advance\@rotdimen by\dp#1%
   \hbox to\@rotdimen{\hskip\ht#1\vbox to\wd#1{\@rotstart{90 rotate}%
   \box#1\vss}\hss}\@rotfinish}
\def\@rotl#1{\@rotdimen=\ht#1\advance\@rotdimen by\dp#1%
   \hbox to\@rotdimen{\vbox to\wd#1{\vskip\wd#1\@rotstart{270 rotate}%
   \box#1\vss}\hss}\@rotfinish}%
\def\@rotu#1{\@rotdimen=\ht#1\advance\@rotdimen by\dp#1%
   \hbox to\wd#1{\hskip\wd#1\vbox to\@rotdimen{\vskip\@rotdimen
   \@rotstart{-1 dup scale}\box#1\vss}\hss}\@rotfinish}%
\def\@rotf#1{\hbox to\wd#1{\hskip\wd#1\@rotstart{-1 1 scale}%
   \box#1\hss}\@rotfinish}%
\def\rotate{\@ifnextchar[{\@rotate}{\@rotate[l]}}
\def\@rotate[#1]#2{\setbox\@rotbox=\hbox{#2}\@nameuse{@rot#1}\@rotbox}

\catcode`\@=12

\begin{document}

\makeatletter \@addtoreset{equation}{section} \makeatother
\renewcommand{\theequation}{\thesection.\arabic{equation}}
\pagestyle{empty}
\setcounter{page}{1} \pagestyle{plain}
\vspace{1.0in}
\rightline{FTUAM-01/18; IFT-UAM/CSIC-01-27}
\vspace{2.5cm}
\setcounter{footnote}{0}

\begin{center}
\large{\bf   
Constraint Supersymmetry Breaking and Non-Perturbative
Effects in String Theory\footnote{To appear in the proceedings of the 
conference ``Recent Developments in General Relativity, Genoa 2000''  
SIGRAV2000, Genoa(Italy), 18-22 September 2000. Based on talks given at
SUSY2K, CERN, Geneva, 26 June-1 July 2000 and at SIGRAV2000}}\\[10mm]
\medskip
C.~Kokorelis\footnote{E-mail : Christos.Kokorelis@uam.es}
\\[1mm]
\small{\em{ Departamento de F\'\i sica Te\'orica C-XI and Instituto de F\'\i sica 
Te\'orica C-XVI
,\\[-0.3em]
Universidad Aut\'onoma de Madrid, Cantoblanco, 28049, Madrid, Spain}}
\end{center}

\begin{center}
\begin{minipage}[h]{14.5cm}
\small{We discuss supersymmetry breaking mechanisms
at the level of low energy ${\cal N }= 1$ effective heterotic
superstring actions that exhibit
$SL(2,Z)_T$ target space modular duality or
$SL(2,Z)_S$
strong-weak coupling duality.
The allowed superpotential forms
use the assumption that the source of non-perturbative effects is
not specified and as a result represent the most
general parametrization of non-perturbative effects.
The minimum values of the limits on the parameters in the
superpotential may correspond
to vacua with vanishing cosmological constant.}
\end{minipage}
\end{center}


One of the biggest problems that heterotic string theory, and its "equivalents", e.g type II,
I,
have to face today
 is the question
of four dimensional ${\cal  N} =1$ space-time supersymmetry breaking. The breaking, due to the
presence of the gravitino, that determines the scale of supersymmetry
breaking,
in the effective action, must be
spontaneous and not explicit. 
Several mechanisms have been used in
recent years to break consistently supersymmetry.
They can be distinguished as to when
they are at work at the string theory level or at the
effective superstring action level.
The first category of mechanisms includes the tree level coordinate dependent
compactification mechanism \cite{CDC}, the
magnetized tori approach \cite{bachas}, the type I brane
breaking \cite{apo1, angel, apo2}, the partial breaking \cite{tayloo}
while the latter category includes approaches that use target space
duality e.g \cite{feto, iba1, bigai, cve} or S-duality \cite{iba2, homou} at the
level of
effective superstring action, related to gaugino condensation \cite{vene},
to constrain
 the allowed superpotential forms.
The main problem in all of the approaches is the creation of 
an appropriate potential for the moduli and the dilaton that 
can fix their vacuum expectation values.
Because
heterotic string theory has target space duality as one of its properties
we can use modular forms to parametrize the unknown non-perturbative
dynamics \cite{cve}, practically to parametrize the unknown 
non-perturbative contributions to the gauge kinetic function $f$.

The purpose of this paper is to reexamine the
issue of constructing superpotentials $W$ that affect supersymmetry
breaking at the level of ${\cal N} = 1$ effective heterotic superstring actions
when the source of non-perturbative effects, is not specified.

 The modification
of the ${\cal N} = 1$ heterotic effective action that we examine in this 
work amounts
to modifying the superpotential when $T$-duality or $S$-duality
non-perturbative effects are included.
Moreover we want to break supersymmetry dynamically, rather
than geometrically, thus we make use of
gaugino condensation in supergravity \cite{tay}.

In string theories physical quantities like masses of
matter, Higgs fields depend on the moduli fields $\Phi$. The latter fields
have flat potential to all orders of string perturbation theory, so their
vacuum expectation values remains undetermined.

In general there are two different approaches in describing
gaugino condensation. These are the effective
lagrangian approach \cite{feto, bigai}, where we can use a
gauge singlet bilinear superfield U as a dynamical degree of 
freedom \cite{tay, la}
and the effective superpotential approach \cite{iba1, cve, luta, lumu, de}.
In the latter formalism
the gauge singlet bilinear superfield is integrated out through
its equation of motion.

The ${\cal N}=1$ effective supergravity
theory coming from superstrings is described by the knowledge
of three functions, the K\"ahler potential $K$, the superpotential $W$
and the gauge kinetic function $f$, that all depend on the moduli fields.

In general duality stabilizes the potentials with local minima at 
the points $T, S = 1, \rho$. 
The minima in the cases considered in \cite{iba2, cve} are either at
the self-dual points giving unbroken space-time supersymmetry in the $T$, $S$
field sector or in the general case supersymmetry breaking minima 
with negative cosmological constant.                 
In the latter case the minima occur at the boundary of the moduli space.

While considerable progess have been made to the
understanding of non-perturbative effects based on D-branes,
the problem of determining the moduli values still remains.
In the absence of such a non-perturbative mechanism, we choose to
break ${\cal N}=1$ local SUSY, via gaugino condensation.
We choose a four dimensional heterotic orbifold \cite{dhvw} with K\"ahler
potential
$K = -3 \log(T + {\bar T})$,
where $T$ the complex structure modulus.

For $(2,2)$ heterotic string compactifications
with ${\cal N} = 1$ supersymmetry
there is at least
one complex modulus $T$ which we denote by $T= R^2 + i b$, where
$R$ is the breathing mode of the six dimensional internal space and $b$ is the internal
axion $\theta$. The $T$-field corresponds, when $T$ large, to the globally
defined $(1, 1)$ K\"ahler form.  
Here we will restrict our study to the simplest (2, 2) models where there is
a single overall modulus, by freezing all other $T$-moduli.

The four dimensional
string effective supergravity
action is invariant to all orders of perturbation theory
under the target space duality transformations
\begin{equation}
T \rightarrow \frac{A T - iB}{i C T + D},\; AD - BC = 1,
\label{enasa}
\end{equation}
since $G = K + \log|W|^2$ has to remain invariant,
while the superpotential transforms wit modular weight $-3$,
$W \rightarrow (i C T + D)^{-3} W$.
Local supersymmetry is spontaneously broken if the auxiliary fields
$h^i=ÝWÝ(^{K/2}(K^i + W^i/W)$ gets non-vanishing
vacuum expectation values. If we choose the superpotential $W$ in the,
factorized, form $W(T,S)= {\Omega(T)K(S)}\eta^{-6}$, $K(S)=e^{-3S/2b}$
that is equivalent to defining the gauge kinetic function
into the form
\begin{eqnarray}
f = S -  \frac{|G_i|}{|G|}(b_a^{N=2}\log[\eta^4(T)(T + {\bar T})]- 
\frac{1}{16\pi^2} Re\{ \partial_T \partial_U h^{(1)}(T, U) - &
\nonumber\\2\log((j(T) -j(U))\}) 
 + (b_a /3) \log|\Omega(T)|^2 + {\cal O}(e^{-S}),
 \label{gauge1}
\end{eqnarray}
where $h^{(1)}$ is the one-loop prepotential\footnote{The one loop
${\cal N} = 2$ four dimensional vector multiplet 
prepotential $h^{(1)}$ was calculated as an ansatz solution to a
differential equation involving one loop corrections to gauge coupling constant
in \cite{hamou}. However, its exact general form for any four
dimensional compactification of the heterotic string was calculated in
\cite{xri}. Higher derivatives of $h^{(1)}$ were calculated in
\cite{lustis}. }, $G_i$ the orders of the subgroup $G$ which 
leaves the i-complex
plane unrotated and we have included the one-loop Green-Schwarz term.
Because $\Omega(T)$ parametrizes our ignorance of
non-perturbative corrections(from target space duality) for the
$T$-modulus, it was assumed \cite{cve} that the $T$-modulus dependent
part of the superpotential takes the form
\begin{eqnarray}
W(T) =  (j-1728)^{m/2} j^{n/3} [\eta(T)]^{2r} {\cal P}(j(T)),\\
W(T) = \Omega(T)[\eta(T)]^{2r} {\cal P}(j(T)),\;\\
\Omega(T)=(j-1728)^{m/2} j^{n/3},
\label{enastria}
\end{eqnarray}
where $m$, $n$ positive integers and $G_6$ , $G_4$ are the Eisenstein
functions of modular weight six
and four and ${\cal P}(j(T))$ an arbitrary polynomial of the
absolutely $SL(2,Z)$ modular invariant $j(T)$. 
Analyzing the supergravity scalar potential corresponding to
(\ref{enastria}) we found negative cosmological constant at its
minimum.
The minima on the cases considered in \cite{iba2, cve} are either at
the self-dual points giving unbroken space-time supersymmetry in the $T$, $S$
field sector respectively or in the general case supersymmetry breaking minima 
with negative cosmological constant.                 
In the latter case the minima occur at the boundary of the moduli space.

In addition, depending on the value of
$(S + {\bar S}) {\partial K(S)}/{\partial S}$ the dilaton minimum
can be at weak or strong coupling \cite{cve}.
However, because $W$ is a section on a flat holomorphic bundle
over the moduli space, the
constraints 
\begin{equation}
n\;mod\;3,m\;mod\;2,
\label{enastesse}
\end{equation}
that have to be included apriori \cite{homou} in
(\ref{enastria}) under the dilatational
transformations $T \rightarrow T + 1$
are missing.  The construction of the superpotentials
that incorporates apriori the constraints (\ref{enastesse})
is performed in \cite{kokostheo}
and
treats not the
$\eta$-invariant as a fundamental quantity, as in (\ref{enastria}),
bur rather the cusp forms $\triangle(T)$ instead.
As a result, by writing down for a moduli field $\Phi$
 the most general weight zero modular
form
\begin{eqnarray}
{j^a(\Phi) (j-1728)^b(\Phi)},
\label{enastesse}
\end{eqnarray}
where $a,b$ integers and $j$ is the $SL(2,Z)$ modular function,
(\ref{enastria}) is generalized\footnote{see for example \cite{kobli}}and the following new $T$-modulus
superpotentials
 are both
allowed, namely
\begin{eqnarray}
W(T) =\frac{{\tilde \Sigma}(T)}{\eta^6(T)}{\cal P}(j),
\label{proto}
\end{eqnarray}
\begin{eqnarray}
W(T) = \frac{1}{\eta^6(T)}\frac{1}{{\tilde \Sigma}(T)}{\cal P}(j).
\label{deytero} 
\end{eqnarray}
Defining the candidate $T$-dual
superpotentials
into the forms (\ref{proto}), (\ref{deytero}) is equivalent  
to demanding that we that we do not allow or do allow poles 
in the upper half plane
respectively. 
This happens because modular functions which are allowed
to have poles in upper half plane are exactly rational functions 
in $j$ (quotients
of polynomials in $j$) whereas modular functions which are not 
allowed to have such
poles
are polynomials in $j$.
Including the dilaton part $K(S)$ in (\ref{proto}), (\ref{deytero})
results in
the construction of the full superpotential.
The superpotentials
    (\ref{proto}), (\ref{deytero})
     can be further constrained by examining stability constraints
     on the scaler potential at the self-dual points $T=1$, $T=\rho=
     e^{i \pi/6}$.
By employing the latter method we derive minimum values on the
$a, b$ parameters,
\begin{eqnarray}
(a ; b)=(1,2,3,..;2,3,..).
\label{con}
\end{eqnarray}
It is worth noticing that the stable minimum values of the $a,b$ parameters
correspond to vacua with vanishing cosmological constant at a generic
point in the moduli space.
The new non-perturbative superpotentials may be used to test the nature of 
non-perturbative
effects in the 4D orbifold constructions of the heterotic string.

\end{document}